\documentclass[useAMS,usenatbib]{mn2e}

 \usepackage{graphicx}
 \usepackage{amsmath}
 \usepackage{amssymb}
 
\newcommand{\mnras}[1]{MNRAS}
\newcommand{\apj}[1]{ApJ}
\newcommand{\apjl}[1]{ApJL}
\newcommand{\apjs}[1]{ApJS}
\newcommand{\aj}[1]{AJ}
\newcommand{\aap}[1]{A\&A}
\newcommand{\aaps}[1]{A\&AS}
\newcommand{\araa}[1]{ARA\&A}
\newcommand{\aapr}[1]{A\&ARv}
\newcommand{\pasp}[1]{PASP}

\newcommand{\msol} { M${_{\sun}}$}
\newcommand{\OC}{$\omega$ Centauri}
\newcommand{\FeH}{[Fe/H]}
\newcommand{\FeIH}{[FeI/H]}
\newcommand{\FeIIH}{[FeII/H]}

\newcommand{\alphaFe}{[$\alpha$/Fe]}

\newcommand{\LaFe}{[LaII/Fe]}
\newcommand{\NaO}{[Na/O]}
\newcommand{\NaFe}{[Na/Fe]}
\newcommand{\OFe}{[O/Fe]}
\newcommand{\SiFe}{[Si/Fe]}
\newcommand{\CaFe}{[Ca/Fe]}
\newcommand{\TiFe}{[Ti/Fe]}
\newcommand{\ScIFe}{[ScI/Fe]}
\newcommand{\ScIIFe}{[ScII/Fe]}
\newcommand{\ScFe}{[Sc/Fe]}
\newcommand{\NiFe}{[Ni/Fe]}
\newcommand{\EuFe}{[EuII/Fe]}
\newcommand{\AlFe}{[Al/Fe]}

\title[Stellar populations in $\omega$ Cen: a multivariate analysis]{Stellar populations in $\omega$ Centauri: a multivariate analysis}
   \author[D. Fraix-Burnet and E. Davoust]{D. Fraix-Burnet$^{1,2}$\thanks{E-mail:
didier.fraix-burnet@obs.ujf-grenoble.fr (DFB); edavoust@irap.omp.eu (ED)}
and 
E. Davoust$^{3}$ \\
$^{1}$ Univ. Grenoble Alpes, IPAG, F-38000 Grenoble, France \\
$^{2}$ CNRS, IPAG, F-38000 Grenoble, France  \\
$^{3}$ IRAP, Universit\'e de Toulouse/CNRS, 14 Avenue Edouard
Belin, F-31400 Toulouse, France 
}
\begin{document} 

 \date{Accepted 1988 December 15. Received 2014 June 23; in original form 2014 June 23}

\pagerange{\pageref{firstpage}--\pageref{lastpage}} \pubyear{2015}

\maketitle

\label{firstpage}

\begin{abstract}
We have performed multivariate statistical analyses of photometric and chemical abundance parameters of three large samples of stars in the globular cluster $\omega$ Centauri.  The statistical analysis of a sample of 735 stars based on seven chemical abundances with the method of Maximum Parsimony (cladistics) yields the most promising results: seven groups are found, distributed along three branches with distinct chemical, spatial and kinematical properties. A progressive chemical evolution can be traced from one group to the next, but also within groups, suggestive of an inhomogeneous chemical enrichment of the initial interstellar matter. The adjustment of stellar evolution models shows that the groups with metallicities \FeH $>$ -1.5 are Helium-enriched, thus presumably of second generation. The spatial concentration of the groups increases with chemical evolution, except for two groups, which stand out in their other properties as well. The amplitude of rotation decreases with chemical evolution, except for two of the three metal-rich groups, which rotate fastest, as predicted by recent hydrodynamical simulations.
The properties of the groups are interpreted in terms of star formation in gas clouds of different origins. 
 In conclusion, our multivariate analysis has shown that metallicity alone cannot segregate the different populations of \OC.

\end{abstract}


%

\section{Introduction}
\label{Introduction}

It is now well established that all globular clusters (or at least all those surveyed in depth) harbour multiple populations of stars. Evidence for this has been found in their spectra, which reveal different abundances of elements, and in their color-magnitude diagrams (hereafter CMD), where distinct main sequences or subgiant branches are visible. While the exact mechanism is still debated, the generally accepted explanation is that these distinct populations are successive generations of stars, the later ones having formed from the ejecta of the earlier generations. Other explanations for these multiple populations, such as mergers of globular clusters, original chemical inhomogeneities or pollution by interacting massive binaries \citep{Bastian2013}, are however not excluded \citep[see e.g. ][ for a recent review]{Gratton2012}.

\OC, the most massive globular cluster of our Galaxy, stands out in this respect. While a modest spread in \FeH\ (0.2 dex or less) has been found in some of the most massive Galactic globular clusters, the spread in \OC\ is more than 1.5 dex \citep[see ][ for a recent update]{Johnson2010}, leading to speculations that it might be the remains of a tidally stripped dwarf galaxy. Being only 5.2 kpc from the Sun, \OC\ has been extensively investigated and the properties of its stellar populations well documented.

Most studies use the CMD to distinguish the different stellar populations of globular clusters, and generally identify stellar sequences with slightly different ages, metallicities and more rarely different helium abundances. \citet{Bedin2004} identify at least two main sequences in the CMD of \OC\ and offer several possible explanations in terms of chemical abundances or pollution by background stars. \citet{Joo2013} identify five different populations of increasing metallicity and helium abundance in the CMD of \OC: a metal-poor one (13.1 Gyr), three of intermediate metallicity (13.0, 12.0 and 11.4 Gyr) and a metal-rich one (11.4 Gyr). Focusing on the subgiant branch of \OC, \citet{Villanova2014} find 6 populations with internal age spreads of over 2 Gyr, and some stars that are merely 4 or 5 Gyr old. 

{Another way of separating different stellar populations is by their chemical abundances, using spectroscopy. Since this is expensive in terms of telescope time, few studies are based on large samples of stars. \citet{Johnson2010}, whose results are used in the present paper, determined abundances of various elements of 855 stars on the red-giant branch of \OC\ from high-resolution (R=20000) spectroscopy. They fit gaussians to the metallicity distribution of these stars, and distinguished four populations in this way: RGB-MP, RGB-Int1, RGB-Int2+3, RGB-a (metal-rich). }

In standard mono-metallic clusters, different generations of stars can also be segregated via their location in the Na-O plane. Using chemical abundances in 19 globular clusters \citet{Carretta2009} derived an empirical criterion for separating the first (or primordial) and second generation of stars. The former, which contains about one third of the stellar populations, is limited to \NaFe $<$ Na$_{min}$ + 0.3. In the latter they distinguished an intermediate component, which contains about 60 to 70\% of the stellar populations and an infrequent extreme one, which is mainly present in more massive clusters.  As will be discussed below, the mixture of stellar populations in \OC\ is too complex for their criterion to be applicable.

There is thus ample evidence for multiple populations in \OC, as well as in other globular clusters, but the methods for distinguishing them are rather empirical, using generally one parameter (the metallicity) or two (those of the CMD).  \citet{Gratton2011} were the first to apply a method of multivariate analysis to separate the stellar populations of globular clusters. They used the data of \citet{Johnson2010}, and applied a k-means algorithm to four parameters of 797 stars: \FeH, \NaO, the average of the abundances of Si, Ca, and Ti, and the abundance of La. They opted for 6 groups (in k-means the number of groups is an input parameter), which they further rearranged into three main groups according to metallicity. They did not consider these three main groups as successive generations, because the Na-O anticorrelation is present in the first two, suggesting the presence of first- and second-generation stars in both. Na and O are correlated in the third main group, unlike in any other globular cluster, but the authors found no evidence for an external origin of these stars.

\citet{Simpson2012} used low-resolution (R=1600) spectroscopic data of 221 red giants in \OC\ to derive their effective temperature, gravity and four abundances, to which they added two abundances from \citet{Johnson2010}. They performed a k-means analysis on six parameters (\FeH\ and the abundances of C, N, O, Na and Ba) under the assumption of four groups. The analysis revealed a low-metallicity group (which they identify as the primordial population), two intermediate-metallicity ones differing in their O and Na abundances (one of them presumably helium-enriched) and a high-metallicity group (in which Na and O are correlated). The two intermediate-metallicity groups are blended in the CMD; when combined, they show the well-known Na-O anticorrelation. The authors identified their low- and high metallicity groups with the corresponding ones of \citet{Gratton2011}. In one of their proposed scenarios the intermediate-metallicity group of high O abundance would be the last one to form, from the gas of the primordial generation, after a gas-sweeping passage through the Galactic plane. In a subsequent study \citep{Simpson2013} they classified 848 stars, 557 of which are new ones, into these four groups through a kind of decision tree. They concluded that formation models of globular clusters and their stellar populations do not provide a consistent picture for the four groups. 

In the present paper we repeat the analyses of \citet{Simpson2012}, \citet{Simpson2013} and \citet{Gratton2011} using a different method of multivariate analysis, that of Maximum Parsimony \citep[cladistics, ][]{jc1,jc2}, which requires no a priori assumption about the number of groups, and which does precisely what the other statistical methods are unable to do: order the groups chronologically. In all this paper, the cladistic analyses were performed with PAUP*4.0b10 \citep{paup} with  the parameters discretized into 30 equal-width bins, which play the role of discrete evolutionary states. More details are given in Appendix~\ref{app:details}. We also apply the k-means algorithm to the data, in order to test the reproducibility of statistical analyses with k-means, and compare the results given by the two methods \citep[see more details on the methods in ][]{Fraix2012}.

\section{Statistical analyses}
\label{statanal}

\begin{table*}
\caption{The 24 parameters and the number of undocumented values from the compilation by \citet{Gratton2011}.}
\label{table:parameters}
\centering
\begin{tabular}{l r l l r}
\hline\hline
\hfil Parameter & \hfil Unknowns && \hfil Parameter & \hfil Unknowns \\
\hline  
V magnitude  &  0 &\phantom{nnn}& \SiFe  & 34 \\
B -- V & 0   && \CaFe  & 1  \\
J magnitude  & 0    && \TiFe  & 32 \\
H magnitude  & 0    && \alphaFe = mean (\SiFe, \CaFe, \TiFe) &  0 \\
K magnitude  & 0    && \NaO   & 14 \\
T$_{\rm eff}$  & 0    && \ScIFe & 441 \\
log(g)  & 0  && \ScIIFe& 100 \\
\FeIH   & 0  && \ScFe = mean(\ScIFe, \ScIIFe) & 37 \\
\FeIIH  & 164 && LaII/Fe& 48 \\
\FeH = mean(\FeIH, \FeIIH)&  0  && \NiFe   &   52 \\
\OFe    &  7 && \EuFe& 721  \\
\NaFe   & 10 && \AlFe &  723  \\
\hline  
\end{tabular}
\end{table*}

\subsection{The Simpson et al (2012, 2013) samples}
\label{Simpson}

We first analyse the sample of 221 stars of \citet{Simpson2012}, using the same six abundance parameters as them. In partitioning analyses like k-means, the number of groups must be given as an input. To estimate the optimum number of groups, we have used several statistical criteria: the Calinski-Harabasz criterion and the Simple Structure Index through the function \textit{cascadeKM} of the \textit{R} package \textit{vegan} \citep{R,vegan}, and the jump technique from \citet{Sugar2003}. Even though these tests are mainly indicative, we find the optimal number to be two, or nine for one of the tests, but never four. As a rationale behind the choice of four groups, \citet{Simpson2012} state that their data are not good enough for dividing their sample into more groups. If we impose four groups, the results of our k-means analysis disagrees with theirs in the number of stars in each group: 9, 52, 60 and 100 in our case, and 24, 49, 62 and 86 in theirs. \citet{Simpson2012} do not provide much details on how they run the k-means algorithm, and it is possible that their study suffers from one well known caveat of the k-means: the result depends a lot on the initial seed. To get rid of this dependence, we used the function \textit{kmeans} of the package \textit{stats} in \textit{R} \citep{R}, which repeats the analysis many times (we chose 1000 repeats) with a different seed and selects the best clustering. No further comparison is possible because the details of their groupings are not provided.

Our cladistic analysis on the sample of 221 stars and six abundances gives six main groups 
which are not in good agreement with the four groups of our k-means clustering described above, but in much better agreement with a k-means analysis with six groups. 

We again do not find any statistical evidence in favour of an optimal number of groups larger than two with the sample of 848 stars from \citet{Simpson2013}. In addition, they do not perform any clustering analysis. They classify by hand 557 stars not present in the first sample by comparing the four parameters they found to characterize best the four groups from their first study. This method has serious limitations, because the larger sample covers a broader domain in the CMD.

\subsection{The Johnson \& Pilachowski sample: four parameters, 797 stars}
\label{stat797}

\citet{Johnson2010} provide eleven abundance parameters for their sample of 855 stars, to which \citet{Gratton2011} added photometry, $T_{\rm eff}$ and log(g)  leading a total of 24 parameters listed in Table~\ref{table:parameters}. If we consider that some of them (\FeH, \ScFe, \alphaFe) should be replaced by mean values to compensate for undocumented values and two (\AlFe\  and \EuFe\ ) should be discarded for lack of data, we end up with 15 parameters usable for a clustering analysis.

To complete the table, we computed the global metallicity $Z$ using log$Z$ = \FeH -- 1.72125 from \citet{Carraro1999}. The range of $Z$ is 0.0004 to 0.003.

\citet{Gratton2011} selected four parameters based on a priori physical arguments for detecting the different generations:

\begin{itemize}
   \item \FeH\  is assumed to be representative of the overall metallicity,
   \item \NaO, which is representative of the location of stars along the Na-O anticorrelation, and is likely to be correlated with the He abundance, although this relation is presumably not linear,
   \item \alphaFe\ is the average of Si, Ca, and Ti, all mainly produced by $\alpha$-capture reactions in massive stars, later exploding as core-collapse SN,
   \item The abundance of La (\LaFe), which is an n-capture element that in the Sun is mainly produced by the s-process.
\end{itemize}

After removal of undocumented stars for these four parameters, the 797-star dataset includes the 221 of \citet{Simpson2012} but \FeH\  is the only parameter in common between the two k-means studies.

We performed the same k-means analysis as \citet{Gratton2011} and found exactly the same distribution in size of the six groups. Unfortunately, the statistical tests prefer two groups, maybe four, but not six. There is thus no statistical justification for the six groups, which means that the algorithm is forced to artificially divide the sample into six groups. \citet{Gratton2011} found that if they choose three groups for the k-means analysis, the differences between the groups are largely dominated by \NaO. Since they do not find this result so informative, they thus increased the number of groups to six. 

The cladistic analysis of the same data set with the same four parameters produces three main groups that include many small ones. The three groups are easily distinguishable in \FeH, \LaFe\  and \alphaFe, and one has a higher average \NaO. Note that, in contrast to the three groups that \citet{Gratton2011} found by k-means clustering, the three groups resulting from our cladistic analysis are not discriminated by \NaO.
Nevertheless, we do not find any really new information with respect to the study by \citet{Gratton2011}. We believe that this is due to the choice of parameters and of number of groups, a choice which is biased by a prioris coming from the physics of stellar evolution, so that the settings of the analysis are not optimal from a statistical point of view.  More precisely, \NaO\  appears in the first component of the Principal Component Analysis, it explains the axis of greatest variance in this sample \citep{Gratton2011}. This does not mean that it is the most discriminant one.

To test the influence of \NaO, we considered Na and O separately, so we use the five parameters \FeH, \NaFe, \OFe, \LaFe\  and \alphaFe\  to find the optimum number of groups for a k-means analysis. The result is more satisfactory since the optimal number is clearly three or four, as compared to only two when one looks for the optimum grouping using the initial 4 parameters. This trend is even stronger if we add log(g).  In other words, the choice of parameters made by \citet{Gratton2011} was probably not the best one for the task. We thus decided to perform a fully objective analysis by selecting the parameters from an entirely statistical point of view. 

\subsection{The Johnson \& Pilachowski sample: seven parameters, 735 stars}
\label{stat735}

We now repeat the clustering analyses on the \citet{Johnson2010} sample with an optimised choice among the 15 usable parameters. In a preliminary step, we checked all the correlations, using all the scatter plots together with a detailed examination of the results from a Principal Component Analysis, looking carefully for the influential parameters within each component. 

After this first step, we kept the V magnitude and discarded H, J and K because the correlations among them are strong and redundancy reasonably present. Also $B-V$ is strongly correlated with $T_{\rm eff}$ and very probably redundant, so that we discard it as well. Following the result of  Sect.~\ref{stat797}, we did not include the ratio \NaO.

We are left with ten parameters: V, log(g), T$_{\rm eff}$, \FeH, \OFe, \NaFe, \ScFe, \NiFe, \LaFe, \alphaFe. However, V,  log(g) and T$_{\rm eff}$ are very much correlated, especially the two first ones, and largely dominate the first Principal Component. In addition, a cladistic analysis using the ten parameters results in a very linear arrangement of the tree according to these three parameters. This indicates a clear redundancy. We thus must discard at least one of them. The two parameters Vmag and T$_{\rm eff}$ have a complicated evolutionary behaviour in the considered region of the HR diagram where the evolutionary tracks (and the isochrones as well) show loops \citep[e.g.][]{Girardi2000,Girardi2004}. These back and forths preclude these parameters from being reliable indicators of a population or a generation. Note that this argument is especially valid for cladistics, but also for partitioning techniques if the goal is to gather objects with the same history.

We finally retain the following seven abundance parameters: \FeH, \OFe, \NaFe, \ScFe, \NiFe, \LaFe, \alphaFe. Some of these parameters have up to 52 undocumented values (Table~\ref{table:parameters}), that is less than 7\%. This is not a problem for cladistics, but k-means, like all distance-based approaches, cannot be used in such a situation. It is possible to replace these missing data for instance by the mean of the each parameter, but we decided to consider a fully documented subset. The final sample has 735 stars, which is slightly fewer than used by \citet{Gratton2011}. 

The optimal number of groups for the k-means analysis is found to be either 2, 3 or large like 35 depending on the algorithm. For comparison with the cladistic result below, we set the number of groups at three or seven. The distribution of stars among the groups is given in Table~\ref{table:confusion}. 

\begin{table*}
\caption{Contingency table comparing the k-means and the cladistic clusterings of the sample with 735 stars described by seven parameters (Sect.~\ref{stat735}). The left part of the table considers the three-group hypothesis, and the right part considers the seven-group hypothesis. }
\label{table:confusion}
\centering
\begin{tabular}{lllllllllllllllll}
\hline\hline
\multicolumn{5}{c}{three groups} && \multicolumn{9}{c}{seven groups} \\
\cline{1-5}
\cline{8-17} 
\noalign{\smallskip}
\hfil Cladistics &\hfil Number&& \multicolumn{3}{c}{k-means} &&\hfil Cladistics &\hfil Number&& \multicolumn{7}{c}{k-means} \\
\hfil group &\hfil of objects &&\multicolumn{3}{c}{} &&\hfil group &\hfil of objects &&\multicolumn{7}{c}{} \\
            &          &&  1     & 2     & 3       &&               &          &&    1  & 2   &   3 &   4  &  5  &  6  &  7    \\
\hline  
OC1  & 230 && 195  & 33  & 2       && OC1    & 230 && 135 & 45  &  3 & 46   & 0  &  0   &  1    \\
OC2     & 265 &&  66   & 190   & 9       && OC2a      & 119  &&  9    &  53 & 4  &  38  & 12 &  0  & 3     \\
            &         &&          &        &           && OC2c      & 99   && 1      & 31 & 54 & 13  & 0   &  0  &  0     \\
            &         &&          &        &           && OC2b & 47   && 0      & 1   &  1  &  45  &  0  &  0  &  0     \\
OC3      &  240 &&  18  & 30    & 192  && OC3a         & 47   && 8      & 0   & 9   &  14  & 10 & 0   & 6       \\
            &         &&          &        &           && OC3c       & 143 && 1      & 1  & 0    & 0     & 61 & 41 & 39     \\
            &         &&          &        &           && OC3b     & 50   &&  0     & 1   &  6   & 0    &  3   &  0   &  40   \\
\hline  
\end{tabular}
\end{table*}

   \begin{figure}
            \includegraphics[width=\linewidth]{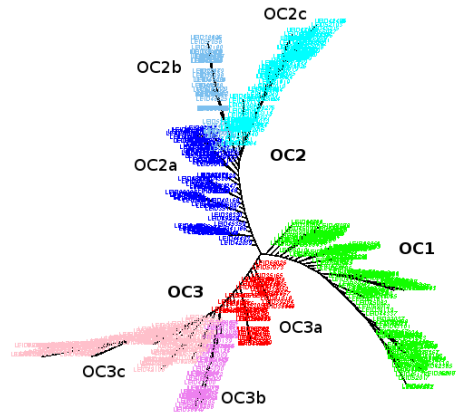}
      \caption{Unrooted tree obtained with cladistics on 735 galaxies described by seven parameters (see Sect.\ref{stat735}). There are three main groups (branches OC1, OC2 and OC3). Groups OC2 and OC3 are each divided into three groups labelled a, b and c (see text). The color codes of the groups will be used throughout the paper. }.
         \label{Fig_treeunrooted}
   \end{figure}

The cladistic analysis of the 735 stars with seven parameters produces a tree that is quite robust (each node occurs in more than 88\% of 9517 equally parsimonious trees). Three main branches, defining three main groups, are identified on Fig.~\ref{Fig_treeunrooted} as OC1, OC2 and OC3. The latter two can be divided into smaller groups: OC2a, OC2b, OC2c, OC3a, OC3b and OC3c, making seven groups altogether. The total number of objects in each group is given in Table~\ref{table:confusion}. Many other sub-branches could define groups but their number of objects would be rather small, and the subsequent interpretation of the tree does not seem to justify this. Note that from a strictly cladistic point of view, the OC2a and OC3a groups may not represent a fully homologous ensemble of objects since they do not include all descendants from the uppermost node. They might be gathered with one of the two other groups (for instance OC2a and OC2b could make a single group from which OC2c diverges), but this cannot be decided without the analyses of the parameters done in the remainder of this paper.

The contingency table (Table~\ref{table:confusion}) provides a detailed comparison between the results of the k-means and cladistic results. For three groups, the k-means result is in fairly good agreement with the cladistic analysis since between 70 and 95 \% of the objects of a group are found in a same group of the other technique. For seven groups, the agreement is poorer. This is not surprising since seven is not found to be an optimal number of groups for the k-means analysis. Strictly speaking, cladistics may identify five groups (see above), but a k-means analysis with this number of groups does not improve the agreement. Maybe a higher number of groups could be pertinent, as suggested by the k-means tests of the optimal number of groups and in agreement with the numerous small subbranches that can be identified on the tree in Fig.~\ref{Fig_treeunrooted}, but the number of stars within each group would be too small for an interesting astrophysical interpretation. In addition, the agreement would certainly not be improved by merely subdividing already discrepant groups. We conclude that the agreement for three main groups is very much satisfactory, and that the k-means approach has some difficulties to find more subtle clustering in this dataset. 

The cladistics result presented in this Section (Fig.~\ref{Fig_treeunrooted}), using 735 stars and seven parameters, is analysed in the remainder of this paper.

\section{Properties of the seven groups}
\label{des:evoltree}

\subsection{Chemical evolution}

Since the cladistic analysis is based on abundance parameters only, one expects the groups to trace the chemical evolution of the stars in \OC. This is indeed what Fig.~\ref{Fig_evoltree} shows, a progressive chemical evolution, as \FeH, \NaO, \AlFe, \alphaFe\ and \LaFe\  increase broadly from the OC1 to the OC2 and OC3 main groups.

   \begin{figure}
            \includegraphics[width=\linewidth]{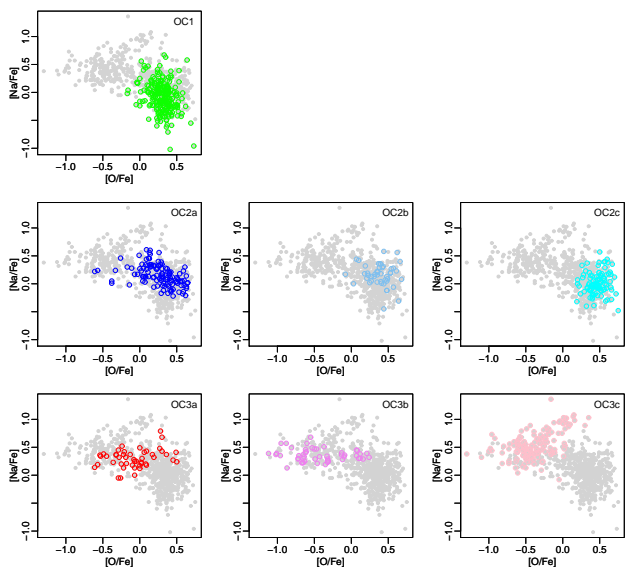}
      \caption{\NaFe\  vs \OFe\  anticorrelation. It is present globally, and to some extent in  OC2a and OC1. The correlation is in OC3c.}
         \label{Fig_NavsO}
   \end{figure}

   \begin{figure}
            \includegraphics[width=\linewidth]{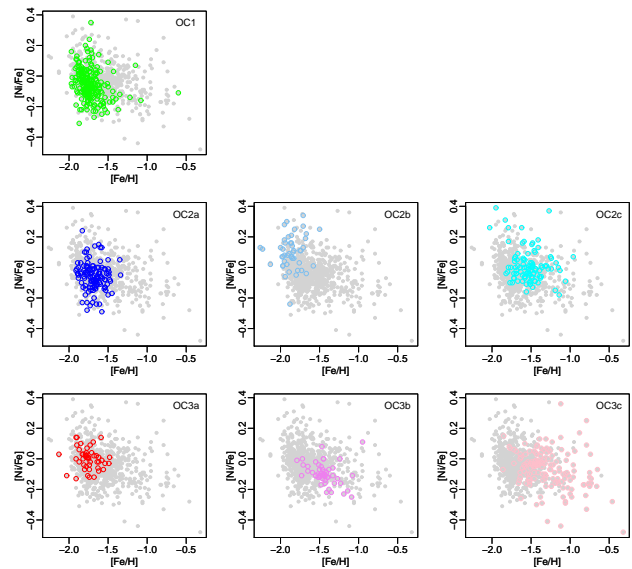}
      \caption{\NiFe\  vs \FeH\  (weak) anticorrelation. It is present in  OC3b and OC3c, and marginally in OC1. The groups OC2a and OC2b are clearly separated in this diagram.}
         \label{Fig_NivsFe}
   \end{figure}

There are distinctive features that explains the division of the three branches into seven groups (Figs.~\ref{Fig_evoltree} and \ref{Fig_boxplots}). Within OC2,  the groups OC2a and OC2b differ in \NiFe, La/Eu and marginally in \alphaFe\  and \AlFe. In OC2b \NiFe\  increases strongly with rank. The group OC2c has lower T$_{\rm eff}$ and log(g), and higher \FeH\  and \alphaFe\  than the two others. It also has brighter K magnitudes.  The group OC2c looks similar to OC3 in that \FeH\  and \alphaFe\  increases with rank, but it has a high \OFe. In other words, the mechanism for producing the Na-O anticorrelation is not at work in OC2c.  

The groups within OC3 have the same  high \NaFe\  and \alphaFe, but differ in some of their abundances, OC3a having lower \FeH, higher \OFe\  and \NiFe\  than the others, and OC3c having the highest \LaFe\  of all.

When examining Fig. A1, one should keep in mind that, cladistics being a most parsimonious approach, the ordering of the stars within branches favors a regular increase of the parameters used for the computation, if they are compatible. This explains why the dispersion is high for the parameters not used in the cladistic analysis (photometry, log(g), T$_{\rm eff}$). Incidentally, in the main group OC1, we note a correlation between V, K, B-V, T$_{\rm eff}$ and log(g) on the one hand, and some abundance parameters, especially the one used to compute the tree, on the other. For instance, as a function of V, we see a decrease of \alphaFe\  and possibly \TiFe,  and an increase of \NaFe, \ScFe, \LaFe. For B-V, the most obvious trend is the increase of \alphaFe. These correlations explain why OC1 is the only group to show a clear trend along its branch.

The Na-O diagram has been much scrutinized in the literature for identifying multiple stellar populations. It is shown in Fig.\ref{Fig_NavsO}. The Na-O anticorrelation, which is present in \OC\ considered as a whole, is absent in individual groups, except in OC2a, and in the form of a slight gradient in the OC1. \NaFe\  is constant in  OC3a and OC3b, and there is a {\it correlation} in OC3c, which makes it identical to the third (metal-rich) group of \citet{Gratton2011} and the fourth (metal-rich) group of \citet{Simpson2012}. In the nomenclature of \citet{Carretta2009},  OC3 can roughly be associated with the extreme population, whereas the two other main groups would be a mixture of primordial and intermediate populations.

Another noticeable anticorrelation among chemical elements in \OC\ is that between \NiFe\  and \FeH. It is weak but significant (correlation coefficient -0.3, p-value=0), as seen on Fig.~\ref{Fig_NivsFe}. Among individual groups it is present in OC3c and OC3b, and weakly in OC1. As mentioned earlier,  OC2a and OC2b are only distinguished by \NiFe. The anticorrelation seems to be present when considering abundances integrated over the whole cluster. \citet{Gratton2004} show such a diagram, but dismiss as deviant the low \NiFe\  in their three metal-rich globular clusters.

\begin{figure}
            \includegraphics[width=\linewidth]{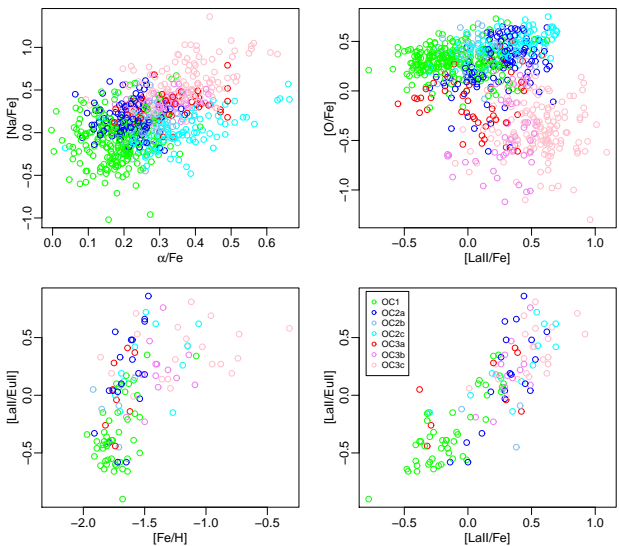}
      \caption{Four projections of the parameter space in which the seven groups occupy distinct areas.}
         \label{Fig_boxplot4}
   \end{figure}

The seven groups occupy distinct locations on three projections of the parameter space: \NaFe\  vs \alphaFe, \LaFe\  vs \OFe\  and La/Eu vs \FeH\  (Fig.~\ref{Fig_boxplot4}). In the first projection, OC1 has low \alphaFe, and there are three parallel diagonal sequences, OC2c, OC2a+OC2b+OC3a+OC3b and OC3c. In the second projection (\LaFe\  vs \OFe), OC1, OC2a+OC2b and OC2c are of increasing \LaFe\  at high \OFe,  OC3a and OC3b are of increasing \LaFe\  as \OFe\  decreases, and OC3c has low \OFe\  and highest \LaFe\ . In the last projection,  OC1 and OC2b have low La/Eu, OC2a shows a large spread, OC3a has intermediate values of La/Eu, and the other groups (OC2c, OC3c and OC3b), which tend to be more metal-rich have a high La/Eu.

\subsection{Comparison with models : log(g)-T$_{\rm eff}$ diagram}
\label{des:g-teff}
 
  \begin{figure}
            \includegraphics[width=\linewidth]{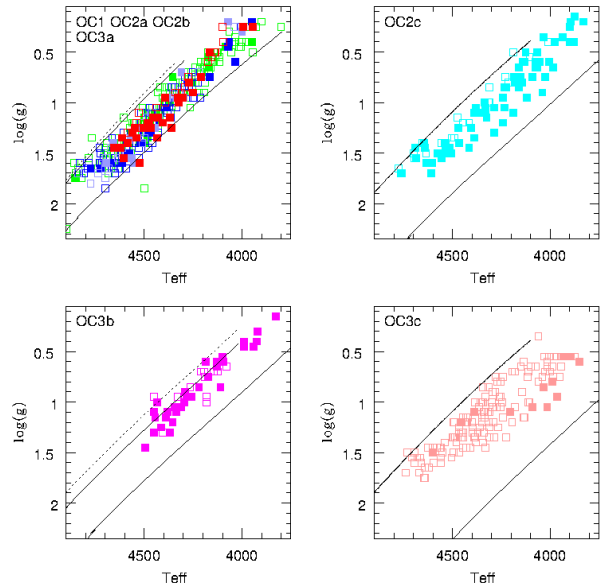}
      \caption{Diagram of effective temperature vs gravity for the different groups. The colors of the groups are the same as in the previous figures. The full symbols correspond to \alphaFe\ $>$ 0.29. The lines are Dartmouth model isochrones for an age of 12.5Gyr. The metallicity and \alphaFe\ values of the different isochrones are those indicated in Table~\ref{table:isochrones}, except in the top left box, where the limits at low metallicity are for \alphaFe\ = 0.2 (solid line) and \alphaFe\ = 0.0 (dotted line). The isochrones are all for a normal He abundance, except in the bottom left box (OC3b), where the limits at low metallicity are for Y = 0.245 (solid line) and Y = 0.4 (dotted line). The He abundances of the high metallicity isochrones for OC2c, OC3b and OC3c are obviously too low.} 
         \label{Fig:isochrones}
   \end{figure}

We now confront the chemical properties of the different groups with stellar evolution models. Our goal is essentially to determine whether Helium enrichment is necessary to explain these properties, and, in that case, which groups are concerned.

To this end we use the Dartmouth stellar evolution models\footnote{http://stellar.dartmouth.edu/models/} \citep{Dotter2008}, which provide isochrones with different abundances of He and $\alpha$-elements. The possible choices of He abundances are Y = 0.245+1.5Z, 0.33 and 0.40. Unfortunately, the higher Y are all for \alphaFe\ = 0. We plot these isochrones in the log(g)-T$_{\rm eff}$ diagram, rather than in the usual color-magnitude diagram, because Teff is independent of distance. In view of the large spread in metallicity in most groups, we plot two isochrones corresponding to the two extreme values (i and f in Table~\ref{table:isochrones}) of \FeH\  and \alphaFe, limits within which the data should in principle be included. 

We stress that the limits in \FeH\  and \alphaFe\  (given in Table~\ref{table:isochrones}) are not free parameters, and that the only free parameters are Y and age. We adopted an age of 12.5 Gyr for all the models; a different value within the permissible range \citep[11.4 to 13.1 Gyr, see ][]{Joo2013} would not displace the isochrone significantly.

The result of the model fitting is shown on Fig.~\ref{Fig:isochrones}. OC1, OC2a, OC2b and OC3a are plotted together in the top left box, because their limits in metallicity are the same, and the differences in limiting \alphaFe\ make little difference in the models, shown as solid (\alphaFe = 0.2) and dotted (\alphaFe\ = 0.0) lines at the low metallicity limit. Most model isochrones are without He overabundances. 
We did plot an isochrone with Y = 0.4 (dotted line in the bottom left box of Fig.~\ref{Fig:isochrones}), which is an extrapolation from the isochrone with \alphaFe\ = 0.0. For the high metallicity limits of OC2c, OC3b and OC3c, we decided that the limiting \alphaFe\ are too high for obtaining a reliable extrapolation. We just note that raising Y lowers the surface gravity of the stars because of the reduced electron scattering opacity, and thus moves the isochrone up in Fig.~\ref{Fig:isochrones} at a given Teff. The consequence is that higher He abundances are required for the limit at high metallicity in groups OC2c, OC3b and OC3c. This confirms previous identifications of He enhancement in this cluster \citep{Norris2004, Piotto2005, Gratton2011, King2012, Joo2013}. The latter authors predicted Y = 0.39-0.41 for the three isochrones with \FeH $>$ -1.55.
\citet{Gratton2011} used the data from \citet{Johnson2010} to estimate Y in their intermediate metallicity population, and found Y = 0.25 for their groups 5 and 1 (average \FeH = -1.636), and Y = 0.347 for their groups 3 and 2a (average \FeH = -1.416). 

This indicates that, at least in \OC, Helium enhancement occurs at metallicities higher than \FeH = -1.5. This is in qualitative agreement with analyses of other globular clusters: 47 Tuc \citep[\FeH = -0.72][]{DiCriscienzo2010}, NGC 2808 \citep[\FeH = -1.14][]{DAntona2005, Piotto2007, Marino2014}, NGC 6388 and NGC 6441 \citep[\FeH = -0.55 and -0.46; e.g. ][]{Caloi2007}, which are all very massive globular clusters with likely He enhancement, and of M3 \citep[\FeH = -1.50][]{Catelan2009} where no He enhancement has been found. This might of course be a coincidence, as so few globular clusters are available for comparison, and one cannot exclude that mass, for example, plays a larger role than metallicity in the He enhancement of stellar populations. There is also indirect evidence that the second generation stars, which seem to be present in all globular clusters, should be He-enhanced \citep[see ][ and references therein]{Gratton2012}. This limit of -1.50 should nevertheless be kept in mind until further notice.

\begin{table}
\caption{The two extreme values (i and f) of the metallicity, \alphaFe\ and Y for the seven groups as used in Fig.~\ref{Fig:isochrones}. Question marks indicate uncertain values, for lack of reliable isochrones.}
\label{table:isochrones}
\centering
\begin{tabular}{l l l l l l l}
\hline\hline
group&\multicolumn{2}{c}{\FeH}&\multicolumn{2}{c}{\alphaFe}&\multicolumn{2}{c}{Y}\\
\hline
&i&f&i&f&i&f\\
\hline
OC1& -2.00& -1.50& 0.0& 0.4& 0.245&0.247\\
OC2a& -2.00& -1.50& 0.1& 0.4& 0.245&0.247\\
OC2b& -2.00& -1.50& 0.2& 0.4& 0.245&0.247\\
OC2c& -1.75& -1.00& 0.2& 0.6& 0.246&0.33?\\
OC3a& -2.00& -1.50& 0.2& 0.5& 0.245&0.247\\
OC3b& -1.50& -1.00& 0.2& 0.4& 0.40 &0.40?\\
OC3c& -1.75& -0.50& 0.2& 0.6& 0.246&0.40?\\
\hline
\end{tabular}
\end{table}

  \begin{figure}
            \includegraphics[width=\linewidth]{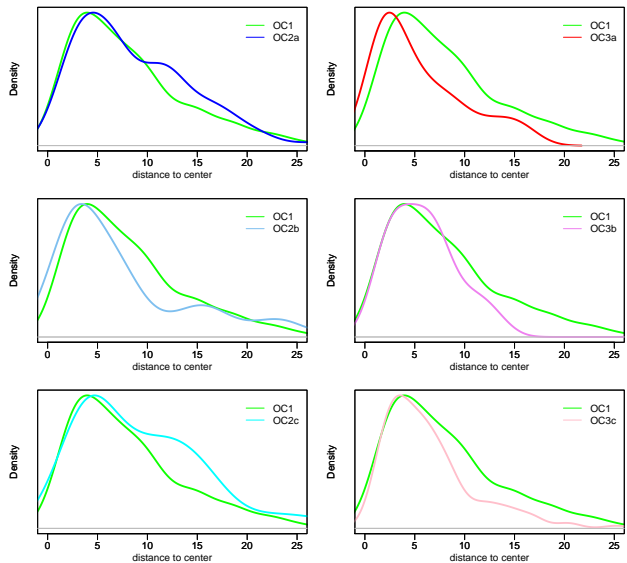}
      \caption{Density distribution of the radial distance (in arcmin) for the different groups. The densities are scaled to the density for OC1 which is shown in all boxes for comparison.}
         \label{Fig_radialdist}
   \end{figure}

\subsection{Radial distribution}
\label{des:radial}

The radial distribution of stars in \OC\ has been studied by many authors, because it can give clues to the star formation history of the cluster. There is a consensus that the intermediate-metallicity RGB stars are more concentrated than the metal-poor ones, but the most metal-rich (\FeH $>$ -0.6) RGB stars, also called RGB-a, are found to be as concentrated as the metal-intermediate ones by some authors \citep{Pancino2000, Johnson2009, Bellini2009}, and the least concentrated by  others \citep[][ their Table 3]{Hilker2000, Castellani2007}.
The situation is more nuanced in our groups, as shown on Fig~\ref{Fig_radialdist}. The least concentrated groups are OC2a and OC2c, followed by OC1 and OC2b. The OC3a, OC3b and OC3c groups are the most concentrated. The contradiction among previous results concerning the metal-rich population is resolved by our clustering of this population into two groups (OC3b and OC3c) of different spatial distributions. 
{
One explanation for the different spatial concentration might be mass segregation, if the groups are dynamically relaxed, which may be the case if the stars have a sufficiently anisotropic velocity distribution (i.e. elongated orbits). But the dynamical relaxation can be questioned and the mass range along the RGB covered by the sample is small.
Following the suggestion of the referee, we propose rather that this group formed where it presently is located, and, for that same reason, is kinematically colder than the others.
}

\citet{Pancino2000} have also found that the metal-poor population is elongated along the east-west (major axis) direction, whereas the two metal-richer populations are elongated in the opposite direction.
\citet{Hilker2000} note that the metal-richest stars are asymmetrically distributed around the center, with an excess toward the south. We do find spatial asymmetries among the groups, but perhaps of marginal significance in view of the small numbers and statistical incompleteness of our sample.
 
\subsection{Kinematics}
\label{des:kin}

The kinematics of the stars in the different groups are likely to provide clues about their formation. Furthermore, if the different groups have different kinematical properties, our clustering results will be validated a posteriori.  We thus collected kinematical data for the stars of our sample. Such data are available from \citet{Suntzeff1996, Mayor1997, Reijns2006, Pancino2007, Sollima2009, Johnson2008}.  We did not use the kinematical data of \citet{vanLoon2007} because of their large quoted uncertainties (8 km/s), even though they would have added 11 stars to the kinematical sample, three of which confirm the large mean rotation of group OC3b (see below). 
We obtained 1433 velocities and uncertainties for 650 stars.  We computed the velocity zero-point of the different samples, and then added 1 km/s to the velocities of \citet{Reijns2006}, and subtracted 2 km/s to those of \citet{Suntzeff1996} and \citet{Pancino2007}.  Retaining only the stars that are in the 735-star sample and  removing 14 velocities which showed a difference of 10 km/s or more with the mean of the others, we took the mean of the measurements for each star, weighted by their uncertainty. This produced a sample of 623 stars with velocities.

We then adjusted a simple rotation curve by a weighted least-squares method to the radial velocities $V_{r}$ of each group, $V_{r} = V_{o}+ 2V_mr_mrsin(\theta - \theta_o)$/($r_m^2$ + $r^2$), where $V_{o}$ = 232.35 km/s is the mean radial velocity of the sample, $r_m$ is the radius at the maximum velocity of rotation $V_{m}$, $\theta$ is the position angle measured from north through east, and $\theta_o$ is the orientation of the axis of rotation with respect to the north-south direction.  We also tried a Toomre rotation curve, which gave comparable results, but with slightly higher dispersions.  We imposed an uncertainty of 0.5 km/s to stars with a single measurement and an uncertainty below that level because such low uncertainties, which strongly affect the fit, seem unrealistic given the spectral resolution.

The kinematical parameters for the different groups are given in Table~\ref{table:kinematics}. N is the number of stars with kinematical data in the (sub)group, and $\sigma$ is the dispersion of velocities around the mean curve. The velocities and velocity dispersions are in km/s, the radii in arcmin, the angles in degrees. The uncertainties on $V_m$ are all less than 0.1 km/s. For group OC3a, the search for the optimal parameters did not converge, because the rotation velocities in that group do not decrease near the origin. For that reason, we fit a rotation curve with constant velocity of rotation in concentric annuli, and, for that group, we list in Table~\ref{table:kinematics} the mean velocity of rotation in different ranges of $r$ (listed in col. 4, instead of $r_m$). For reference, we note that the mean peak rotation velocity of \OC, determined with larger samples, is 8 km/s, and is reached at about 8 arcmin \citep{Merritt1997, vandeVen2006}.

The results show that the groups indeed have different kinematics. The groups OC1, OC2a and OC2c rotate like the bulk of the stars;  OC3a, OC3b and OC2b rotate very fast in the inner region (where  OC3a and OC3b are mostly located); OC3c shows very little rotation, if any. When significant, rotation is roughly around the minor axis, which makes an accretion scenario unprobable, except perhaps for OC3a, which does not seem to be in dynamical equilibrium, as its rotation pattern is independent of the mass distribution in the cluster.

We are not the first to find that the subpopulations of \OC\ have different kinematics. \citet{Norris1997} and \citet{vandeVen2006} found that the metal-rich component, which is spatially more centrally concentrated, shows no evidence of systematic rotation.  These results were contradicted by \citet{Pancino2007} and \citet{vanLoon2007}.  \citet{Ferraro2002} found that the metal-rich, so-called "RGB-a" component shows a different proper-motion signature from the metal-poor and metal-intermediate components. This has been disputed by \citet{vandeVen2006, Bellini2009, Anderson2010}. Incidentally, the latter found that the proper-motion dispersion increases with decreasing stellar mass, suggesting that the core of \OC\ is in the process of establishing energy equipartition. The explanation for these conflicting results is that one cannot segregate the different populations of \OC\ on the basis of metallicity alone.

\begin{table}
\caption{The kinematical parameters of the seven groups, assuming a variable velocity of rotation, except for OC3a}
\label{table:kinematics}
\centering
\begin{tabular}{l|rrrr|rrrr}
\hline\hline
\hfil Group & \hfil N & \hfil $V_m$ & \hfil $r_m$ & \hfil $\theta_o$&\hfill $\sigma$\\
\hline
OC1&194& 7.4&11.4&322& 8.06\\
OC2a&104& 7.1& 8.3&354& 7.53\\
OC2b& 38&13.3& 4.0&328& 8.93\\
OC2c& 91& 8.2& 8.2&347& 9.48\\
OC1+OC2a+OC2c&389&7.4&10.2&337& 8.29\\
OC3a& 8&18.4& r$<$3.5 &80&13.3\\
OC3a&12&16.6& 3.5$\le$r$<$7&12&5.24\\
OC3a&10& 4.7& 7$\le$r&306&6.04\\
OC3b& 46&17.3& 2.0&328& 8.49\\
OC3c&120& 3.6& 3.3&  6& 7.76\\
\hline
\end{tabular}
\end{table}

\section{Discussion}
\label{discussion}

\subsection{Primordial and later generations}

We now attempt to relate our groups to populations and/or successive generations of stars. and, to this end, look for evidence of external pollution in the chemical properties of the groups. 

The OC1 group is obviously a primordial population. It has the lowest mean value of most chemical elements, such as \NaFe, \AlFe, \NiFe, \LaFe, La/Eu, \alphaFe, and a high \OFe. At variance with the other groups, there is no marked evolution of the abundances with rank within the group, except perhaps Ni and Eu. {The main evolution with rank is a decrease in the V and K magnitudes and in T$_{eff}$ and log(g), which were not used in the statistical analysis. }
The group has a spatial distribution that is close to the mean one in the cluster, and rotates as expected from the mass distribution of the cluster \citep{vandeVen2006}, suggesting that it is dynamically relaxed. Finally, this group has 31\% of the stars, which is about the expected percentage for the primordial population (see Sect.~\ref{Introduction}).

We find three hints of external pollution in the chemical properties of the OC2 group: the presence of the Na-O anticorrelation and a large spread in [LaII/EuII] in the OC2a group, and a significantly larger \NiFe\  in the OC2b group. 
The Na-O anticorrelation is generally attributed to pollution from H-burning at high temperature, in either fast rotating massive stars \citep{Decressin2007} or intermediate-mass asymptotic giant branch (AGB) stars \citep{DAntona2007}.
Ni is a heavier element than Fe, thus produced in supernova nucleosynthesis. 
\citet{Tsujimoto2003} attribute the increasing La/Eu in \OC\ to pollution by AGB stars after gas removal by supernovae of type II (hereafter SN II) at the end of a first episode of star formation. 
We note however that these two groups have the same O abundance as OC1, and that they are not markedly He-enriched, indicating that they formed from gas that was not much polluted.

The OC2c group belongs to the OC2 group, but, like the OC3c group, differs markedly from the other groups of its main group, and these two outstanding groups should be discussed together. Both are significantly He-enriched (see Table~\ref{table:isochrones}) and have high \alphaFe; this makes them likely to belong to a secondary population. The difference is in \AlFe, low in OC2c and high in OC3c. Both groups are also spatially extended, while  OC2a and OC3a  are centrally concentrated. 

The group OC3 is significantly Al-richer than the other groups.  Al-rich stars must be second-generation stars because of the high temperatures necessary for the MgAl reaction \citep{Langer1993}, which cannot be reached in low-mass stars. Furthermore, since He is the main outcome of hot H burning, these polluted stars must also be He-enriched. However, only OC3c and OC3b are significantly He-enriched (see Table~\ref{table:isochrones}). The group OC3a, is very similar to OC2a and OC2b, except for \AlFe, which is the highest of all groups, and \OFe, which is low enough to make it part of an extreme secondary population. 
An Na-O correlation, rather than anticorrelation, is present in OC3c. It can be explained by advocating super-AGB stars of masses between 6.5 and 8\msol, whose ejecta contribute to the formation of new stars about 120 to 200 Myr after the first generation \citep{DAntona2011}.

In summary, there is evidence for increasing external pollution in the stars of  OC2a and OC3a, but the outstanding properties of OC2c and OC3c, as well as the properties of OC3a, intermediate between those of OC2a and OC3a, do not call for a simple division of the sample into three populations, let alone successive generations, of stars, as done by \citet{Carretta2009} for monometallic globular clusters.

\subsection{Origin of the enriched gas}

The next step is to use the spatial and kinematical properties of the groups, as well as their chemical ones, to identify possible origins for the enriched gas that produced the secondary populations. This gas was either expelled by winds from earlier stars or by supernovae.

Gas ejected from stars with strong winds will disperse in interstellar space. Such stars can be either fast rotating massive stars or massive AGBs; the gas will thus be produced a few tens or hundreds of Myr after the onset of the star formation event. Some of this gas may form stars when it is shocked during a passage of the cluster through the galactic plane \citep{Ostriker1972}, before the gas density in the disc is high enough for ram pressure stripping to occur \citep{Tsujimoto2003}. Since the period of rotation of \OC\ is 120 Myr \citep{Dinescu1999}, this plane crossing is likely to happen every 60 Myr. \citet{Leon2000} have found evidence for disc shocking in \OC, in the form of a tidal tail aligned with the tidal field gradient. The stars formed in this way will show little mean rotation, if any. The remainder of the ejected gas will likely fall radially into the strong potential well of the cluster, generating stars that will move in very elongated orbits.  The gas that originates in stars with slower winds will presumably follow the stars in their orbits until close to the pericenter, where it collides with other gas clouds and settles into circular orbits, along which stars will form. 

This scenario, which is inspired from the one proposed by \citet{Eggen1962} to interpret the properties of first- and second-generation dwarf stars in the Galaxy, can explain OC3. The group OC3c (chemically evolved and spatially extended) shows little or no mean rotation, while OC3a and OC3b (chemically evolved and centrally concentrated) have the fastest mean rotation. This faster rotation has also been predicted by \citet{Bekki2010, Bekki2011}; his hydrodynamical simulations show that the gaseous ejecta from AGB stars produce stars with different kinematics from the ones of the primordial population, namely a higher mean rotation and a lower mean velocity dispersion. In any case,  OC3a and OC3b have to be fairly young, because this kinematical difference with the other groups cannot survive many Gigayears of dynamical friction. The core relaxation time (at a radius of 2.5 arcmin) of \OC\ is 4 Gyr \citep[][ 2010 edition\footnote{http://physwww.physics.mcmaster.ca/~harris/mwgc.dat}]{Harris1996}, suggesting ages of that order. There are alternative explanations for faster rotation, either the gravo-gyro instability, which affects massive stars more than others \citep{Kim2004}, or a central binary black-hole, which produces "suprathermal" stars at a few core radii \citep{Mapelli2005}, but such mechanisms do not favor metal-rich stars over metal-poor ones. 

Supernova explosions are the most common source of enriched gas in the Galaxy. The progenitors of SN II are massive stars, whose lifetime is shorter than 100 Myr, and can be as short as a few Myr for the most massive ones. The progenitors of SN 1a are binary stars; their peak of occurrence is 40-50 Myr \citep{Matteucci2001} or 100 Myr \citep{Kobayashi2009}. However their frequency in globular clusters is probably much lower than that of SN II \citep{Voss2012, Washabaugh2013}. Both types are more likely to occur in the central region of the cluster, where massive and binary stars are preferentially found.  SN II produce O and $\alpha$-elements, but little Fe, while SN 1a produce mainly $^{56}$Ni (which decays to form $^{56}$Fe), and other isotopes of Ni in lower proportions, but not $\alpha$-elements, so that \alphaFe\  actually decreases \citep[][, their Table 3]{Iwamoto1999}.  SN 1a enrichment has been found in a few stars of the  metal-rich "RGB-a" branch of \OC\ \citep{Pancino2011}. According to \citet{Marcolini2007}, SN 1a occur in small pockets of gas before being mixed with the interstellar medium, at variance with SN II, which pollute the interstellar medium uniformly. Stars formed in these pockets are thus likely to retain the angular momentum of the progenitor gas clouds. This mechanism could explain the formation of OC2b, although its \alphaFe\ is not lower than average.

The problem with enrichment by SN II is that these might be too energetic events for the gas to be retained by the cluster \citep[e.g. ][]{Tsujimoto2003, Baumgardt2008}, although radiative losses may prevent this from happening in a very massive cluster \citep{Marcolini2007}. SN events might not be an important source of enriched gas, thus perhaps not at the origin of the groups beyond OC1, but they can very well have triggered star formation by compressing and heating the gas enriched by other means.

\subsection{Formation scenario for the groups}

The formation scenario for the different groups of stars in \OC\ must call on discrete events. This is required by the distinct isochrones found in the CMD of the cluster (see Sect.~\ref{Introduction}), which lend credibility to our clustering of the stars into groups with distinct properties. Another requirement of the scenario is to take into account the large metallicity spread in the primordial and secondary populations, which is unique among galactic globular clusters. The current favored explanation is that the cluster is the remnant of a captured dwarf galaxy \citep[e.g. ][]{Dinescu1999, Bekki2006, Marcolini2007}. 
{Our analysis has not enabled us to provide an explanation for the large metallicity spread in the primordial group. }

In the primordial population (OC1), the red-giant branch is the narrowest and the abundances are the least dispersed and the most constant with rank, compared to the other groups. These are properties of a population formed over a short period of time in a single event of star formation (a single generation, making a true isochrone) in a homogeneous environment (very little spread in the evolutionary track).  

Judging from the chemical properties of the different groups and from their tracks on the CMD, OC2 and OC3 must have formed successively from progressively enriched gas. We propose that OC2 formed from early SN explosions, before AGB pollution had time to enrich the gas, except perhaps for OC2a, in which there is some evidence for the Na-O anticorrelation. The group OC2b is enriched in O and Ni, which indicates pollution from SN 1a ejecta. This group is also more concentrated than OC2a and OC2c. In order to explain the high rotational velocity of this group, we have to assume that the SN 1a pockets formed near the center in or after the re-collapse phase at the end of the SN II events, thus acquiring kinetic energy. The group OC2c formed over the full extent of the cluster during early galactic plane crossings. 

The group OC3, which has the highest \AlFe\  and \NaFe\ and the lowest \OFe, formed last, in an interstellar medium O-depleted and Na-enriched, and also possibly He-enriched, by relatively massive AGB stars. The groups OC3a and OC3b formed by the mechanism proposed by \citet{Bekki2010, Bekki2011} in the central regions, with higher rotation velocities, while OC3c formed like OC2c.

We realise that many questions are left unanswered by this scenario but it outlines possible reasons for the existence of seven distinct groups of stars.

\section{Conclusion}

We have performed a multivariate statistical analysis of the chemical properties of star samples in \OC. {At variance with previous studies of the stellar populations of this cluster, we made no a priori assumptions about the number of groups}. In addition, our technique for obtaining the grouping is based on the relationships between the stars and not on their similarities, an approach that is suited to find populations and generations. Using the full range of known chemical properties, rather than just metal abundances, allows one to perform a clustering into groups that is much more efficient at revealing the distinctive properties of the stellar populations in a sample that at first sight shows a bewildering array of chemical, spatial and kinematical properties. 

We found seven groups, which can be gathered into three main groups, and identified OC1 as the primordial population. The other groups show progressive chemical enrichment, while retaining distinct properties, for which we provide a tentative interpretation in terms of locus and mechanism of formation. Group OC2b formed near the center from SNIa ejecta and OC2a formed after AGB stars had chemically enriched the interstellar medium. OC3a and OC3b formed later, from the ejecta of stars with slow winds, OC3a being still dynamically unrelaxed. OC3c formed like OC2c, during Galactic plane crossings, but at a later stage of formation of the Galaxy.  

Our proposed clustering into groups lifts the contradictions between the results of earlier studies relative to the spatial distribution and kinematics of the different stellar populations, which cannot be distinguished by metallicity alone. It also proposes directions in which future investigations could proceed to explain the uncovered properties. In particular, it remains to be shown whether the Na-O correlation found in OC3c can be produced by star formation during Galactic plane crossings, and whether the He-rich stars presently observed in globular clusters should on average be more massive than if they had formed with a normal He abundance.

\section*{Acknowledgments}

We thank Aaron Dotter for comments on the Dartmouth models, and  Andr\'e Maeder for his comments on a preliminary version of this paper. We warmly thank the referee, Grabriele Gratton, for his very construitive comments that helped us improve the paper. 


\begin{thebibliography}{74}
\expandafter\ifx\csname natexlab\endcsname\relax\def\natexlab#1{#1}\fi

\bibitem[{{Anderson} \& {van der Marel}(2010)}]{Anderson2010}
{Anderson} J., {van der Marel} R.~P., 2010, \apj, 710, 1032

\bibitem[{{Bastian} {et~al.}(2013){Bastian}, {Lamers}, {de Mink}, {Longmore},
  {Goodwin}, \& {Gieles}}]{Bastian2013}
{Bastian} N., {Lamers} H.~J.~G.~L.~M., {de Mink} S.~E., {Longmore} S.~N.,
  {Goodwin} S.~P., {Gieles} M., 2013, \mnras, 436, 2398

\bibitem[{{Baumgardt} {et~al.}(2008){Baumgardt}, {Kroupa}, \&
  {Parmentier}}]{Baumgardt2008}
{Baumgardt} H., {Kroupa} P., {Parmentier} G., 2008, \mnras, 384, 1231

\bibitem[{{Bedin} {et~al.}(2004){Bedin}, {Piotto}, {Anderson}, {Cassisi},
  {King}, {Momany}, \& {Carraro}}]{Bedin2004}
{Bedin} L.~R., {Piotto} G., {Anderson} J., {Cassisi} S., {King} I.~R., {Momany}
  Y., {Carraro} G., 2004, \apjl, 605, L125

\bibitem[{{Bekki}(2010)}]{Bekki2010}
{Bekki} K., 2010, \apjl, 724, L99

\bibitem[{{Bekki}(2011)}]{Bekki2011}
---, 2011, \mnras, 412, 2241

\bibitem[{{Bekki} \& {Norris}(2006)}]{Bekki2006}
{Bekki} K., {Norris} J.~E., 2006, \apjl, 637, L109

\bibitem[{{Bellini} {et~al.}(2009){Bellini}, {Piotto}, {Bedin}, {King},
  {Anderson}, {Milone}, \& {Momany}}]{Bellini2009}
{Bellini} A., {Piotto} G., {Bedin} L.~R., {King} I.~R., {Anderson} J., {Milone}
  A.~P., {Momany} Y., 2009, \aap, 507, 1393

\bibitem[{{Caloi} \& {D'Antona}(2007)}]{Caloi2007}
{Caloi} V., {D'Antona} F., 2007, \aap, 463, 949

\bibitem[{{Carraro} {et~al.}(1999){Carraro}, {Girardi}, \&
  {Chiosi}}]{Carraro1999}
{Carraro} G., {Girardi} L., {Chiosi} C., 1999, \mnras, 309, 430

\bibitem[{{Carretta} {et~al.}(2009){Carretta}, {Bragaglia}, {Gratton},
  {Lucatello}, {Catanzaro}, {Leone}, {Bellazzini}, {Claudi}, {D'Orazi},
  {Momany}, {Ortolani}, {Pancino}, {Piotto}, {Recio-Blanco}, \&
  {Sabbi}}]{Carretta2009}
{Carretta} E., {Bragaglia} A., {Gratton} R.~G., {Lucatello} S., {Catanzaro} G.,
  {Leone} F., {Bellazzini} M., {Claudi} R., {D'Orazi} V., {Momany} Y.,
  {Ortolani} S., {Pancino} E., {Piotto} G., {Recio-Blanco} A., {Sabbi} E.,
  2009, \aap, 505, 117

\bibitem[{{Castellani} {et~al.}(2007){Castellani}, {Calamida}, {Bono},
  {Stetson}, {Freyhammer}, {Degl'Innocenti}, {Moroni}, {Monelli}, {Corsi},
  {Nonino}, {Buonanno}, {Caputo}, {Castellani}, {Dall'Ora}, {Del Principe},
  {Ferraro}, {Iannicola}, {Piersimoni}, {Pulone}, \& {Vuerli}}]{Castellani2007}
{Castellani} V., {Calamida} A., {Bono} G., {Stetson} P.~B., {Freyhammer} L.~M.,
  {Degl'Innocenti} S., {Moroni} P.~P., {Monelli} M., {Corsi} C.~E., {Nonino}
  M., {Buonanno} R., {Caputo} F., {Castellani} M., {Dall'Ora} M., {Del
  Principe} M., {Ferraro} I., {Iannicola} G., {Piersimoni} A.~M., {Pulone} L.,
  {Vuerli} C., 2007, \apj, 663, 1021

\bibitem[{{Catelan} {et~al.}(2009){Catelan}, {Grundahl}, {Sweigart},
  {Valcarce}, \& {Cort{\'e}s}}]{Catelan2009}
{Catelan} M., {Grundahl} F., {Sweigart} A.~V., {Valcarce} A.~A.~R.,
  {Cort{\'e}s} C., 2009, \apjl, 695, L97

\bibitem[{{D'Antona} {et~al.}(2005){D'Antona}, {Bellazzini}, {Caloi}, {Pecci},
  {Galleti}, \& {Rood}}]{DAntona2005}
{D'Antona} F., {Bellazzini} M., {Caloi} V., {Pecci} F.~F., {Galleti} S., {Rood}
  R.~T., 2005, \apj, 631, 868

\bibitem[{{D'Antona} {et~al.}(2011){D'Antona}, {D'Ercole}, {Marino}, {Milone},
  {Ventura}, \& {Vesperini}}]{DAntona2011}
{D'Antona} F., {D'Ercole} A., {Marino} A.~F., {Milone} A.~P., {Ventura} P.,
  {Vesperini} E., 2011, \apj, 736, 5

\bibitem[{{D'Antona} \& {Ventura}(2007)}]{DAntona2007}
{D'Antona} F., {Ventura} P., 2007, \mnras, 379, 1431

\bibitem[{{Decressin} {et~al.}(2007){Decressin}, {Meynet}, {Charbonnel},
  {Prantzos}, \& {Ekstr{\"o}m}}]{Decressin2007}
{Decressin} T., {Meynet} G., {Charbonnel} C., {Prantzos} N., {Ekstr{\"o}m} S.,
  2007, \aap, 464, 1029

\bibitem[{{di Criscienzo} {et~al.}(2010){di Criscienzo}, {Ventura}, {D'Antona},
  {Milone}, \& {Piotto}}]{DiCriscienzo2010}
{di Criscienzo} M., {Ventura} P., {D'Antona} F., {Milone} A., {Piotto} G.,
  2010, \mnras, 408, 999

\bibitem[{{Dinescu} {et~al.}(1999){Dinescu}, {Girard}, \& {van
  Altena}}]{Dinescu1999}
{Dinescu} D.~I., {Girard} T.~M., {van Altena} W.~F., 1999, \aj, 117, 1792

\bibitem[{{Dotter} {et~al.}(2008){Dotter}, {Chaboyer}, {Jevremovi{\'c}},
  {Kostov}, {Baron}, \& {Ferguson}}]{Dotter2008}
{Dotter} A., {Chaboyer} B., {Jevremovi{\'c}} D., {Kostov} V., {Baron} E.,
  {Ferguson} J.~W., 2008, \apjs, 178, 89

\bibitem[{{Eggen} {et~al.}(1962){Eggen}, {Lynden-Bell}, \&
  {Sandage}}]{Eggen1962}
{Eggen} O.~J., {Lynden-Bell} D., {Sandage} A.~R., 1962, \apj, 136, 748

\bibitem[{{Ferraro} {et~al.}(2002){Ferraro}, {Bellazzini}, \&
  {Pancino}}]{Ferraro2002}
{Ferraro} F.~R., {Bellazzini} M., {Pancino} E., 2002, \apjl, 573, L95

\bibitem[{{Fraix-Burnet}(2009)}]{DFB09}
{Fraix-Burnet} D., 2009, Evolutionary Biology Concept, Modeling, and
  Application, {Pontarotti} P., ed., Biomedical and Life Sciences, Springer
  Berlin Heidelberg, pp. 363--378

\bibitem[{{Fraix-Burnet} {et~al.}(2012){Fraix-Burnet}, {Chattopadhyay},
  {Chattopadhyay}, {Davoust}, \& {Thuillard}}]{Fraix2012}
{Fraix-Burnet} D., {Chattopadhyay} T., {Chattopadhyay} A.~K., {Davoust} E.,
  {Thuillard} M., 2012, \aap, 545, A80

\bibitem[{{Fraix-Burnet} {et~al.}(2006{\natexlab{a}}){Fraix-Burnet}, {C}holer,
  \& {D}ouzery}]{FCD06}
{Fraix-Burnet} D., {C}holer P., {D}ouzery E., 2006{\natexlab{a}}, {A}stronomy
  and {A}strophysics, 455, 845

\bibitem[{{Fraix-Burnet} {et~al.}(2006{\natexlab{b}}){Fraix-Burnet}, {C}holer,
  {D}ouzery, \& {V}erhamme}]{jc1}
{Fraix-Burnet} D., {C}holer P., {D}ouzery E., {V}erhamme A.,
  2006{\natexlab{b}}, {J}. {C}lassification, 23, 31, 16 pages, 6 figures

\bibitem[{{Fraix-Burnet} {et~al.}(2009){Fraix-Burnet}, {D}avoust, \&
  {C}harbonnel}]{FDC09}
{Fraix-Burnet} D., {D}avoust E., {C}harbonnel C., 2009, MNRAS, 398, 1706

\bibitem[{{Fraix-Burnet} {et~al.}(2006{\natexlab{c}}){Fraix-Burnet}, {D}ouzery,
  {C}holer, \& {V}erhamme}]{jc2}
{Fraix-Burnet} D., {D}ouzery E., {C}holer P., {V}erhamme A.,
  2006{\natexlab{c}}, {J}.{C}lassification, 23, 57, 14 pages, 4 figures

\bibitem[{Girardi {et~al.}(2000)Girardi, Bressan, Bertelli, \&
  Chiosi}]{Girardi2000}
Girardi L., Bressan A., Bertelli G., Chiosi C., 2000, \aaps, 141, 371

\bibitem[{Girardi {et~al.}(2004)Girardi, Grebel, Odenkirchen, \&
  Chiosi}]{Girardi2004}
Girardi L., Grebel E.~K., Odenkirchen M., Chiosi C., 2004, \aap, 422, 205

\bibitem[{{Gratton} {et~al.}(2004){Gratton}, {Sneden}, \&
  {Carretta}}]{Gratton2004}
{Gratton} R., {Sneden} C., {Carretta} E., 2004, \araa, 42, 385

\bibitem[{{Gratton} {et~al.}(2012){Gratton}, {Carretta}, \&
  {Bragaglia}}]{Gratton2012}
{Gratton} R.~G., {Carretta} E., {Bragaglia} A., 2012, \aapr, 20, 50

\bibitem[{{Gratton} {et~al.}(2011){Gratton}, {Johnson}, {Lucatello}, {D'Orazi
  D'Orazi}, \& {Pilachowski}}]{Gratton2011}
{Gratton} R.~G., {Johnson} C.~I., {Lucatello} S., {D'Orazi D'Orazi} V.,
  {Pilachowski} C., 2011, \aap, 534, A72

\bibitem[{{Harris}(1996)}]{Harris1996}
{Harris} W.~E., 1996, \aj, 112, 1487

\bibitem[{{Hilker} \& {Richtler}(2000)}]{Hilker2000}
{Hilker} M., {Richtler} T., 2000, \aap, 362, 895

\bibitem[{{Iwamoto} {et~al.}(1999){Iwamoto}, {Brachwitz}, {Nomoto},
  {Kishimoto}, {Umeda}, {Hix}, \& {Thielemann}}]{Iwamoto1999}
{Iwamoto} K., {Brachwitz} F., {Nomoto} K., {Kishimoto} N., {Umeda} H., {Hix}
  W.~R., {Thielemann} F.-K., 1999, \apjs, 125, 439

\bibitem[{{Johnson} \& {Pilachowski}(2010)}]{Johnson2010}
{Johnson} C.~I., {Pilachowski} C.~A., 2010, \apj, 722, 1373

\bibitem[{{Johnson} {et~al.}(2009){Johnson}, {Pilachowski}, {Michael Rich}, \&
  {Fulbright}}]{Johnson2009}
{Johnson} C.~I., {Pilachowski} C.~A., {Michael Rich} R., {Fulbright} J.~P.,
  2009, \apj, 698, 2048

\bibitem[{{Johnson} {et~al.}(2008){Johnson}, {Pilachowski}, {Simmerer}, \&
  {Schwenk}}]{Johnson2008}
{Johnson} C.~I., {Pilachowski} C.~A., {Simmerer} J., {Schwenk} D., 2008, \apj,
  681, 1505

\bibitem[{{Joo} \& {Lee}(2013)}]{Joo2013}
{Joo} S.-J., {Lee} Y.-W., 2013, \apj, 762, 36

\bibitem[{{Kim} {et~al.}(2004){Kim}, {Lee}, \& {Spurzem}}]{Kim2004}
{Kim} E., {Lee} H.~M., {Spurzem} R., 2004, \mnras, 351, 220

\bibitem[{{King} {et~al.}(2012){King}, {Bedin}, {Cassisi}, {Milone}, {Bellini},
  {Piotto}, {Anderson}, {Pietrinferni}, \& {Cordier}}]{King2012}
{King} I.~R., {Bedin} L.~R., {Cassisi} S., {Milone} A.~P., {Bellini} A.,
  {Piotto} G., {Anderson} J., {Pietrinferni} A., {Cordier} D., 2012, \aj, 144,
  5

\bibitem[{{Kobayashi} \& {Nomoto}(2009)}]{Kobayashi2009}
{Kobayashi} C., {Nomoto} K., 2009, \apj, 707, 1466

\bibitem[{{Langer} {et~al.}(1993){Langer}, {Hoffman}, \& {Sneden}}]{Langer1993}
{Langer} G.~E., {Hoffman} R., {Sneden} C., 1993, \pasp, 105, 301

\bibitem[{{Leon} {et~al.}(2000){Leon}, {Meylan}, \& {Combes}}]{Leon2000}
{Leon} S., {Meylan} G., {Combes} F., 2000, \aap, 359, 907

\bibitem[{{Mapelli} {et~al.}(2005){Mapelli}, {Colpi}, {Possenti}, \&
  {Sigurdsson}}]{Mapelli2005}
{Mapelli} M., {Colpi} M., {Possenti} A., {Sigurdsson} S., 2005, \mnras, 364,
  1315

\bibitem[{{Marcolini} {et~al.}(2007){Marcolini}, {Sollima}, {D'Ercole},
  {Gibson}, \& {Ferraro}}]{Marcolini2007}
{Marcolini} A., {Sollima} A., {D'Ercole} A., {Gibson} B.~K., {Ferraro} F.~R.,
  2007, \mnras, 382, 443

\bibitem[{{Marino} {et~al.}(2014){Marino}, {Milone}, {Przybilla}, {Bergemann},
  {Lind}, {Asplund}, {Cassisi}, {Catelan}, {Casagrande}, {Valcarce}, {Bedin},
  {Cort{\'e}s}, {D'Antona}, {Jerjen}, {Piotto}, {Schlesinger}, {Zoccali}, \&
  {Angeloni}}]{Marino2014}
{Marino} A.~F., {Milone} A.~P., {Przybilla} N., {Bergemann} M., {Lind} K.,
  {Asplund} M., {Cassisi} S., {Catelan} M., {Casagrande} L., {Valcarce}
  A.~A.~R., {Bedin} L.~R., {Cort{\'e}s} C., {D'Antona} F., {Jerjen} H.,
  {Piotto} G., {Schlesinger} K., {Zoccali} M., {Angeloni} R., 2014, \mnras,
  437, 1609

\bibitem[{{Matteucci} \& {Recchi}(2001)}]{Matteucci2001}
{Matteucci} F., {Recchi} S., 2001, \apj, 558, 351

\bibitem[{{Mayor} {et~al.}(1997){Mayor}, {Meylan}, {Udry}, {Duquennoy},
  {Andersen}, {Nordstrom}, {Imbert}, {Maurice}, {Prevot}, {Ardeberg}, \&
  {Lindgren}}]{Mayor1997}
{Mayor} M., {Meylan} G., {Udry} S., {Duquennoy} A., {Andersen} J., {Nordstrom}
  B., {Imbert} M., {Maurice} E., {Prevot} L., {Ardeberg} A., {Lindgren} H.,
  1997, \aj, 114, 1087

\bibitem[{{Merritt} {et~al.}(1997){Merritt}, {Meylan}, \&
  {Mayor}}]{Merritt1997}
{Merritt} D., {Meylan} G., {Mayor} M., 1997, \aj, 114, 1074

\bibitem[{{Norris}(2004)}]{Norris2004}
{Norris} J.~E., 2004, \apjl, 612, L25

\bibitem[{{Norris} {et~al.}(1997){Norris}, {Freeman}, {Mayor}, \&
  {Seitzer}}]{Norris1997}
{Norris} J.~E., {Freeman} K.~C., {Mayor} M., {Seitzer} P., 1997, \apjl, 487,
  L187

\bibitem[{Oksanen {et~al.}(2013)Oksanen, Blanchet, Kindt, Legendre, Minchin,
  O'Hara, Simpson, Solymos, Stevens, \& Wagner}]{vegan}
Oksanen J., Blanchet F.~G., Kindt R., Legendre P., Minchin P.~R., O'Hara R.~B.,
  Simpson G.~L., Solymos P., Stevens M. H.~H., Wagner H., 2013, vegan:
  Community Ecology Package. R package version 2.0-10

\bibitem[{{Ostriker} {et~al.}(1972){Ostriker}, {Spitzer}, \&
  {Chevalier}}]{Ostriker1972}
{Ostriker} J.~P., {Spitzer} Jr. L., {Chevalier} R.~A., 1972, \apjl, 176, L51

\bibitem[{{Pancino} {et~al.}(2000){Pancino}, {Ferraro}, {Bellazzini}, {Piotto},
  \& {Zoccali}}]{Pancino2000}
{Pancino} E., {Ferraro} F.~R., {Bellazzini} M., {Piotto} G., {Zoccali} M.,
  2000, \apjl, 534, L83

\bibitem[{{Pancino} {et~al.}(2007){Pancino}, {Galfo}, {Ferraro}, \&
  {Bellazzini}}]{Pancino2007}
{Pancino} E., {Galfo} A., {Ferraro} F.~R., {Bellazzini} M., 2007, \apjl, 661,
  L155

\bibitem[{{Pancino} {et~al.}(2011){Pancino}, {Mucciarelli}, {Bonifacio},
  {Monaco}, \& {Sbordone}}]{Pancino2011}
{Pancino} E., {Mucciarelli} A., {Bonifacio} P., {Monaco} L., {Sbordone} L.,
  2011, \aap, 534, A53

\bibitem[{{Piotto} {et~al.}(2007){Piotto}, {Bedin}, {Anderson}, {King},
  {Cassisi}, {Milone}, {Villanova}, {Pietrinferni}, \& {Renzini}}]{Piotto2007}
{Piotto} G., {Bedin} L.~R., {Anderson} J., {King} I.~R., {Cassisi} S., {Milone}
  A.~P., {Villanova} S., {Pietrinferni} A., {Renzini} A., 2007, \apjl, 661, L53

\bibitem[{{Piotto} {et~al.}(2005){Piotto}, {Villanova}, {Bedin}, {Gratton},
  {Cassisi}, {Momany}, {Recio-Blanco}, {Lucatello}, {Anderson}, {King},
  {Pietrinferni}, \& {Carraro}}]{Piotto2005}
{Piotto} G., {Villanova} S., {Bedin} L.~R., {Gratton} R., {Cassisi} S.,
  {Momany} Y., {Recio-Blanco} A., {Lucatello} S., {Anderson} J., {King} I.~R.,
  {Pietrinferni} A., {Carraro} G., 2005, \apj, 621, 777

\bibitem[{{R Core Team}(2014)}]{R}
{R Core Team}, 2014, R: A Language and Environment for Statistical Computing. R
  Foundation for Statistical Computing, Vienna, Austria

\bibitem[{{Reijns} {et~al.}(2006){Reijns}, {Seitzer}, {Arnold}, {Freeman},
  {Ingerson}, {van den Bosch}, {van de Ven}, \& {de Zeeuw}}]{Reijns2006}
{Reijns} R.~A., {Seitzer} P., {Arnold} R., {Freeman} K.~C., {Ingerson} T., {van
  den Bosch} R.~C.~E., {van de Ven} G., {de Zeeuw} P.~T., 2006, \aap, 445, 503

\bibitem[{{Simpson} \& {Cottrell}(2013)}]{Simpson2013}
{Simpson} J.~D., {Cottrell} P.~L., 2013, \mnras, 433, 1892

\bibitem[{{Simpson} {et~al.}(2012){Simpson}, {Cottrell}, \&
  {Worley}}]{Simpson2012}
{Simpson} J.~D., {Cottrell} P.~L., {Worley} C.~C., 2012, \mnras, 427, 1153

\bibitem[{{Sollima} {et~al.}(2009){Sollima}, {Bellazzini}, {Smart}, {Correnti},
  {Pancino}, {Ferraro}, \& {Romano}}]{Sollima2009}
{Sollima} A., {Bellazzini} M., {Smart} R.~L., {Correnti} M., {Pancino} E.,
  {Ferraro} F.~R., {Romano} D., 2009, \mnras, 396, 2183

\bibitem[{Sugar \& James(2003)}]{Sugar2003}
Sugar C.~A., James G.~M., 2003, J. American Statistical Assoc., 98, 750

\bibitem[{{Suntzeff} \& {Kraft}(1996)}]{Suntzeff1996}
{Suntzeff} N.~B., {Kraft} R.~P., 1996, \aj, 111, 1913

\bibitem[{Swofford(2003)}]{paup}
Swofford D.~L., 2003, Paup*: Phylogenetic analysis using parsimony (*and other
  methods)

\bibitem[{{Tsujimoto} \& {Shigeyama}(2003)}]{Tsujimoto2003}
{Tsujimoto} T., {Shigeyama} T., 2003, \apj, 590, 803

\bibitem[{{van de Ven} {et~al.}(2006){van de Ven}, {van den Bosch}, {Verolme},
  \& {de Zeeuw}}]{vandeVen2006}
{van de Ven} G., {van den Bosch} R.~C.~E., {Verolme} E.~K., {de Zeeuw} P.~T.,
  2006, \aap, 445, 513

\bibitem[{{van Loon} {et~al.}(2007){van Loon}, {van Leeuwen}, {Smalley},
  {Smith}, {Lyons}, {McDonald}, \& {Boyer}}]{vanLoon2007}
{van Loon} J.~T., {van Leeuwen} F., {Smalley} B., {Smith} A.~W., {Lyons} N.~A.,
  {McDonald} I., {Boyer} M.~L., 2007, \mnras, 382, 1353

\bibitem[{{Villanova} {et~al.}(2014){Villanova}, {Geisler}, {Gratton}, \&
  {Cassisi}}]{Villanova2014}
{Villanova} S., {Geisler} D., {Gratton} R.~G., {Cassisi} S., 2014, \apj, 791,
  107

\bibitem[{{Voss} \& {Nelemans}(2012)}]{Voss2012}
{Voss} R., {Nelemans} G., 2012, \aap, 539, A77

\bibitem[{{Washabaugh} \& {Bregman}(2013)}]{Washabaugh2013}
{Washabaugh} P.~C., {Bregman} J.~N., 2013, \apj, 762, 1

\end{thebibliography}

\bsp

\appendix

\section{Maximum Parsimony (cladistic) Analysis}
\label{app:details}

The cladistic analysis in astrophysics (astrocladistic) has been presented in detail in several papers\footnote{see also http://astrocladistics.org} \citep{FCD06,jc1,jc2,DFB09,FDC09}. We refer the reader to these references for a complete description of the method used here.

Phylogenetic systematics relies on synapomorphies, which are derived character states, to infer common ancestry relationships. Such characters may be viewed as evolutionary novelties appearing in a particular lineage. It is assumed that two closely relative objects share derived characters, which presumably originate in their common ancestor.

An important advantage of phylogenetic systematics is to avoid the grouping of objects based on similarities due to evolutionary convergences or reversals. The success of a cladistic analysis depends on the behaviour of the parameters. In particular, it is sensitive to redundancies, incompatibilities, and especially to homoplasies (reversals, parallel and convergent evolutions). 
The use of the parsimony principle  minimises the number of homoplasies (see below).

We have discarded Vmag and Teff since they are certainly homoplasies (see Sect.~\ref{stat735}). The seven chemical parameters used for the analysis can reasonably be considered as synapomorphies since these abundances are generated within stars before being ejected to form new stars. Hence they are specific to the ancestors that gave birth to a given population. Nevertheless, we cannot not guarantee them to be perfect synapomorphies because of measurement uncertainties and of our lack of a complete and thorough understanding of the chemical evolution of the gas within stars and during ejection.

The values for each parameter were discretized into 30 equal-width bins representing supposedly evolutionary states. This choice of 30 bins is justified by a fair representation of the diversity and of the continuous nature of the data, as well as a good stability of the analysis in the sense that the result does not depend on the number of bins.

We imposed all the parameters to be ordered, i.e. changes between two adjacent states are more probable than between distant ones, independently from the sense of the change. We believe that this is a fair measure of the cost of evolution for continuous parameters.

We adopted the popular parsimony criterion, which selects the most parsimonious tree among all possible arrangements because it represents the simplest evolutionary scenario compatible with the input data. This optimal tree corresponds to the minimum number of changes of states for all parameters that occur along the paths between all objects. This number is unique to each tree. 

The maximum parsimony searches were performed using the heuristic algorithm implemented in the PAUP*4.0b10 \citep{paup} package. Heuristic methods do not explore the parameter space of all possible tree arrangements, which would take a prohibitive computer time for hundreds of objects, but try to find the minima. They cannot guarantee finding the absolute most parsimonious trees
but generally require far less computer time while being quite effective. 

Fig~\ref{Fig_treeunrooted} is a majority-rule consensus tree reflecting the most common features in all equally most parsimonious trees.

\begin{figure*}
            \includegraphics[width=\linewidth]{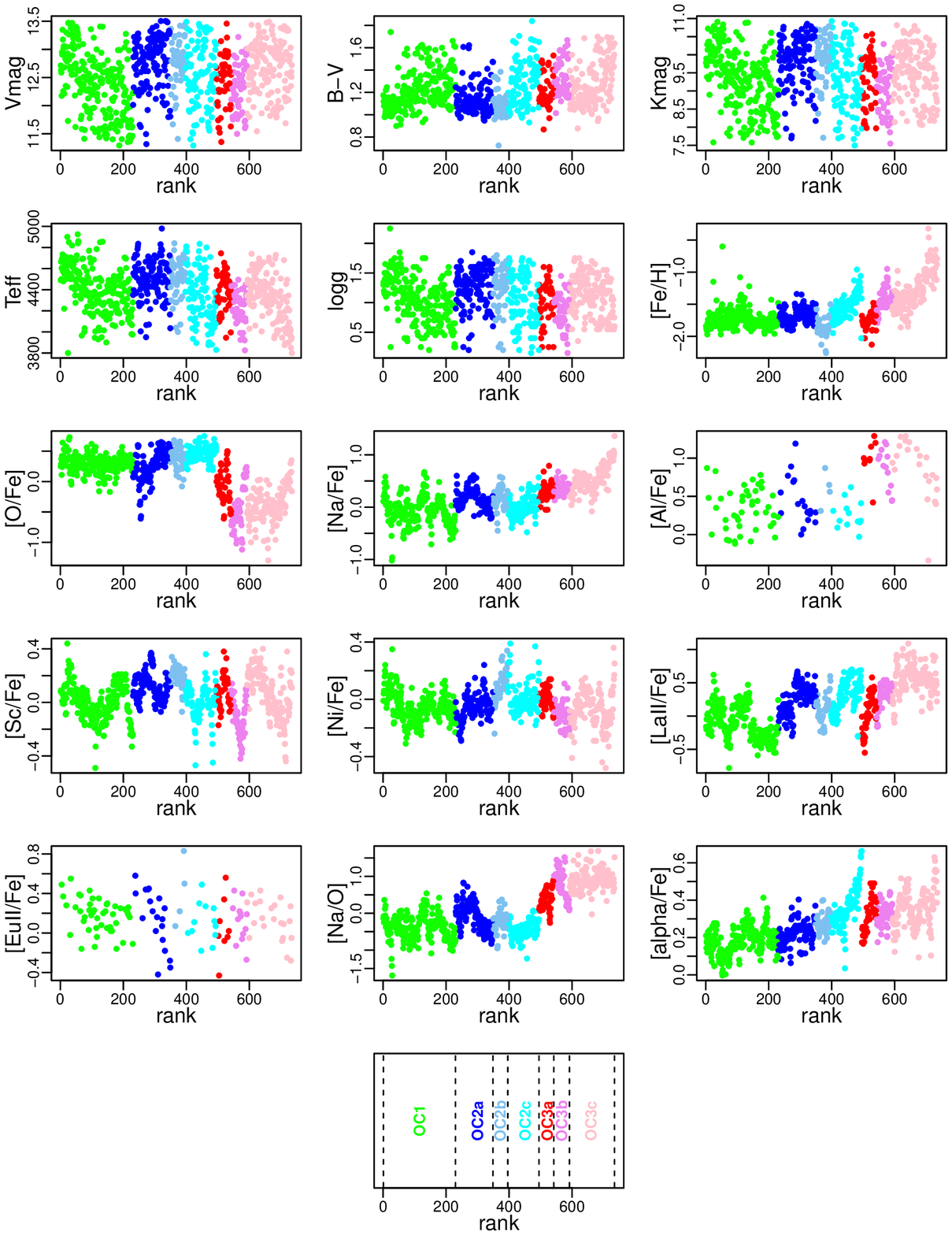}
      \caption{Evolution of each parameter as a function of the rank along the tree. The bottom plot gives the locations of the groups.}
         \label{Fig_evoltree}
   \end{figure*}

\begin{figure*}
            \includegraphics[width=0.9\linewidth]{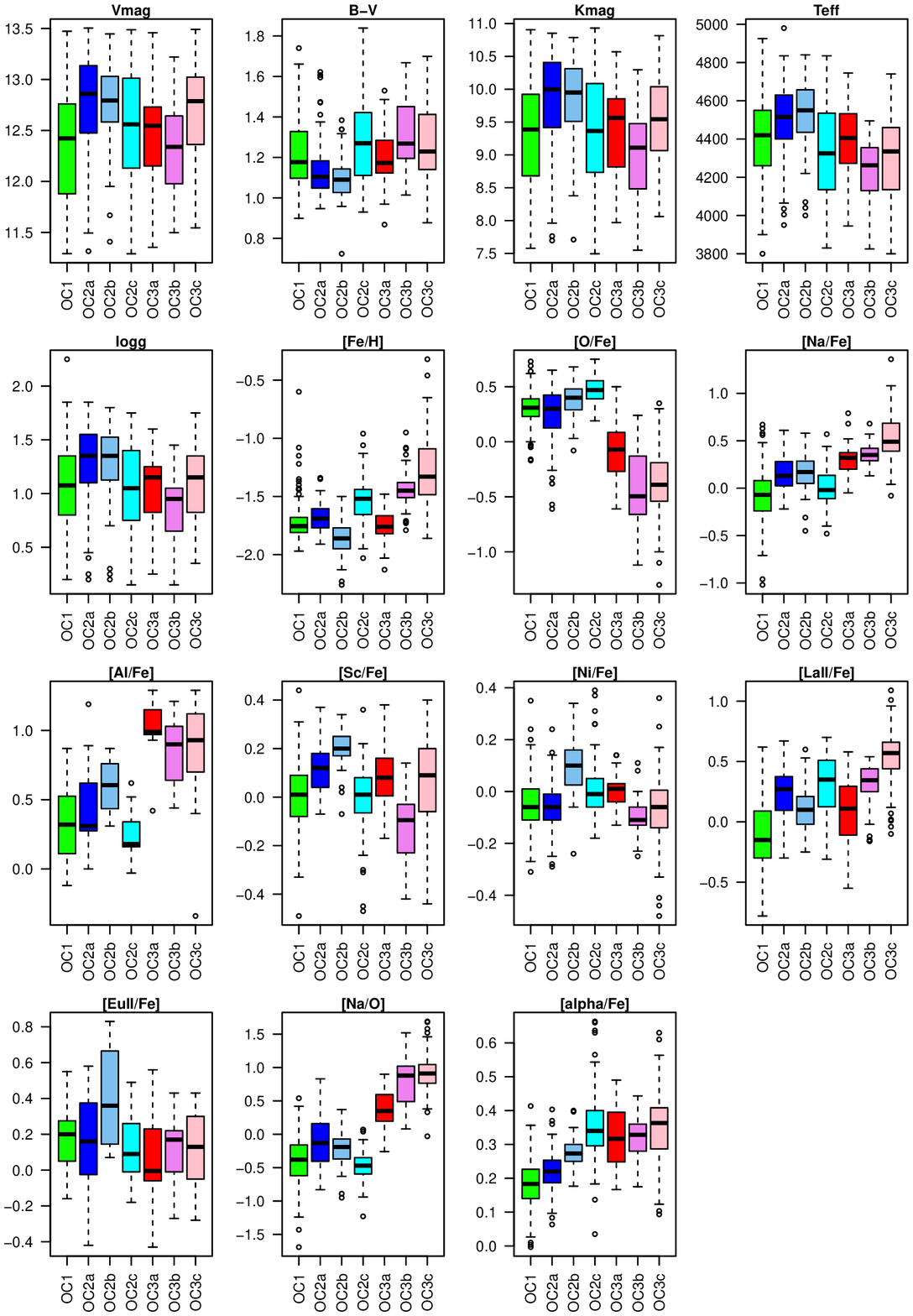}
      \caption{Boxplots}
         \label{Fig_boxplots}
   \end{figure*}

The careful examination of the resulting tree must be performed to check a posteriori the behaviour of the parameters. Firstly, the robustness of the tree is a good indicator of the quality of the characters (see below). Secondly, the evolution of the parameters along the tree can use for instance  Fig.~\ref{Fig_evoltree} or Fig.~\ref{Fig_boxplots}. Note that this should ideally be done with the binned parameters. Other analyses can then be repeated after removal of suspicious parameters. For instance, \ScFe\ and \NiFe\ seem to be too much variable on Fig.~\ref{Fig_evoltree}. An analysis without these two parameters still produce the three main groups. 

The robustness of the tree was assessed in the following way:
\begin{enumerate}
   \item using several analyses modifying sligthly the set of parameters. This is the philosophy of the bootstrap technique, which would be too much time consuming due to the large number of objects. In all cases, the result did not change much in the sense that the three main groups can be identified quite easily.
   \item considering the quality of the consensus tree. A consensus is a summary of all the equally parsimonious trees found by the software. We found each node to be present at least in 89\% of the 9517 equally parsimonious trees, most being in 99 or 100\% cases.
\end{enumerate}

\section{Complementary figures}

\label{lastpage}

\end{document}